\documentclass[reprint, aps, pre,nofootinbib]{revtex4-1}
\usepackage[T1]{fontenc}
\usepackage[latin9]{inputenc}
\usepackage{color}
\usepackage{array}
\usepackage{float}
\usepackage{multirow}
\usepackage{amsmath}
\usepackage{amssymb}
\usepackage{graphicx}
\usepackage{esint}

\makeatletter

\pdfpageheight\paperheight
\pdfpagewidth\paperwidth

\newcommand*\LyXZeroWidthSpace{\hspace{0pt}}
\providecommand{\tabularnewline}{\\}

\makeatother

\usepackage{babel}
\begin{document}
\title{Normal and anomalous diffusion in a bouncing ball over an irregular
surface}
\author{Ana Laura Boscolo}
\author{Valdir Barbosa da Silva Junior}
\author{Luiz Antonio Barreiro}
\email{luiz.a.barreiro@unesp.br}

\address{S\~ao Paulo State University (Unesp), Institute of Geosciences and Exact
Sciences, ~\linebreak{}
Physics Department, CEP 13506-900, Rio Claro, S\~ao Paulo, Brazil}
\date{\today}
\begin{abstract}
The problem of a bouncing ball on a non-planar surface is investigated.
We discovered that surface undulation adds a horizontal component
to the impact force, which acquires a random character. Some aspects
of Brownian motion are found in the horizontal distribution of the
particle. On the x-axis, normal and super diffusion are observed.
For the probability density's functional form, a scaling hypothesis
is presented. 
\end{abstract}
\maketitle

\section{Introduction}

Diffusion is a common natural phenomenon and generally occurs when
a system moves toward the equilibrium state \citep{Ma1985}. Many
domains employ the notion of diffusion, including physics (particle
diffusion), chemistry, biology, sociology, economics, and finance
\citep{Berryman1977,Shlesinger1986,Yu2003}. They all represent the
fundamental concept of diffusion, which asserts that a substance or
collection expands away from a point or location where that material
or collection is more concentrated. In a diffusion process in a set
of moving elements - energy, linear momentum, atoms, molecules, cells,
animals, etc - each element performs a random trajectory. As a result
of this highly irregular individual movement, the ensemble diffuses.
Many non-linear systems also present a diffusive behavior in your
phase space. Modeling such a dynamic system has become one of the
most challenging subjects among scientists. The modeling helps to
understand in many cases how the system evolves in time \citep{Lichtenberg1992,Strogatz,Strogatz1994}.

On a macroscopic level, the average collective behavior, in contrast
to the microscopic individual movement, shows great regularity and
follows well-defined dynamic laws. The non-linear dynamic formulation
of these transport phenomena, as well as the diffusion equation, are
two ways to describe the diffusion phenomena. The form of the temporal
dependence of the mean squared distance (MSD), $\left\langle x^{2}\right\rangle \propto t^{2\mu}$,
or, equivalently, of the variance, allows classifying the type of
diffusion. For $\mu=1/2$ we have the usual or normal diffusion, which
can be described by Fick's laws. Otherwise, we have an anomalous diffusion
(or non-Fickian diffusion). When $\mu>1/2$ the case is classified
as superdiffusive \citep{Geisel,Szymanski} and for $\mu<1/2$ we
have the subdiffusive case \citep{Saxton2001,Massignan2014}. Indeed,
a wide diversity of systems presents a non-linear growth of the mean
squared displacement.

\textcolor{black}{In this work, we explore the diffusive behavior
that occurs in a free-falling particle colliding with a non-planar
surface. Compared to a flat surface, on which the falling particles
maintain their velocity in the horizontal direction, a non-planar
surface introduces changes in the horizontal component of velocity
after each collision. This creates a spread in the absolute value
of the horizontal component of velocity as well as in its signal.
Thus, in section II we study the dynamics of the model, in which the
equations of motion are established, and how the iterative process
takes place. Some special points are explored in \ref{subsec:Periodic-points},
for which no diffusion is observed. In section III, the randomness
of the horizontal component of the collision force is studied. Also,
the diffusion in the signal of the horizontal component of velocity
and its relation to the random walk problem are explored. Section
IV is devoted to discussing the behavior of the mean square displacement
in relation to the initial collision point and the Probability Distribution
Function (PDF) numerically and analytically. In section V, the conclusions
and final considerations about the problem addressed are presented.}

\section{The Model}

\begin{figure}[H]
\centering{}\includegraphics[scale=0.4]{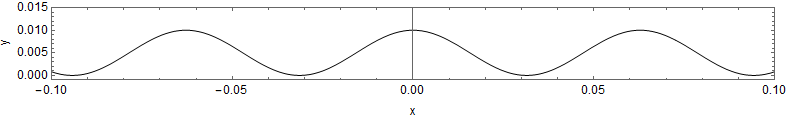}\caption{Graph obtained from equations (\ref{xyspt}) using the parameters
$\alpha=0.01$ and $\beta=0.001$.}
\label{FigFloor}
\end{figure}

We now discuss how to construct the equations of the mapping that
describe the dynamics of the particles. The model under study consists
of an ensemble of non-interacting classical particles of mass $m$
traveling in the presence of a constant gravitational field $\boldsymbol{g}$
and colliding with a non-flat ground via elastic collisions. The parametric
equations that describe the ground are: 
\begin{equation}
\begin{split}x(p) & =\alpha\,p\\
y(p) & =\beta\left[1+\textrm{cos}\left(p\right)\right],
\end{split}
\label{xyspt}
\end{equation}
The figure \ref{FigFloor} shows an example with $\alpha=0.01$ and
$\beta=0.001$.
Here it is worth noting that if the $\beta$ parameter is null then
the floor becomes flat, recovering the traditional Bouncer model \citep{Livorati}
with a static floor. However, different from the traditional Bouncer
model, if $\beta\neq0$, the particles gain an extra degree of freedom,
with movement in the $x$-direction too. Also, as in the Bouncer model,
the action of the constant gravitational field $g$ is responsible
for the return mechanism of the particle for the next collision with
the floor. The conservation of energy during the collision is controlled
by a parameter which is called the \textit{coefficient of restitution}
and it is denoted by $\gamma$. For $\gamma=1$ the conservative dynamics
is observed. However, if $0<\gamma<1$ we found a dissipative behavior.

\subsection{The Map}

We now explore the time evolution of particles, determining the coordinates
of the collision points and their respective velocities. The dynamic
evolution of the particle can be described by the Newton's equation
of motion
\begin{equation}
m\frac{d\boldsymbol{v}}{dt}=\boldsymbol{F}_{grav}+\boldsymbol{F}_{col},\label{Newton1}
\end{equation}
where $\boldsymbol{F}_{grav}=m\boldsymbol{g}$ is the gravitational
force acting on the particle and $\boldsymbol{F}_{col}$ represents
the instantaneous force of collision with the ground. We will assume
that the collision force only changes the velocity component orthogonal
to the surface. It is also an acceptable assumption that during the
collision process the force $\boldsymbol{F}_{col}$ has an extremely
rapid variation. 

\begin{figure}[H]
\begin{centering}
\includegraphics[scale=0.5]{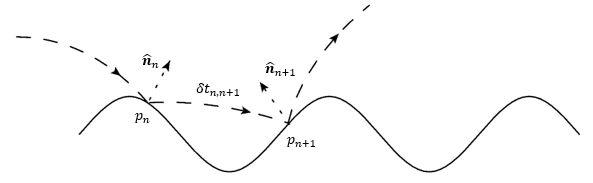}
\par\end{centering}
\caption{Schematic drawing of the trajectory of a particle, with its collision
points and the respective normal vectors.}
\label{FigCollision}
\end{figure}

A typical path taken by the particles is shown in the figure \ref{FigCollision}.
After the $n$th collision at the point defined by the parameter $p_{n}$,
the particle travels in the gravitational field until it collides
at the point $p_{n+1}$. This journey takes a $\delta t_{n,n+1}$
time and continues incessantly if no dissipation is taken into account.
The normal vectors at each collision point are also shown. The unit
normal and tangent vectors at the point $p_{n}$ can be written in
terms of the Cartesian vectors as 
\begin{equation}
\hat{\boldsymbol{n}}_{n}=\frac{\left(-\lambda_{n}\:\boldsymbol{i}+\boldsymbol{j}\right)}{\sqrt{1+\lambda_{n}^{2}}}\;\textrm{and}\;\hat{\boldsymbol{t}}_{n}=\frac{\left(\boldsymbol{i}+\lambda_{n}\:\boldsymbol{j}\right)}{\sqrt{1+\lambda_{n}^{2}}}\label{normal1}
\end{equation}
where $\lambda_{n}$ is the local inclination of the ground, which
for the functions in (\ref{xyspt}), is given by
\begin{equation}
\lambda_{n}=\frac{(dy/dp)_{p_{n}}}{(dx/dp)_{p_{n}}}=-\frac{\beta}{\alpha}\textrm{sin}\left(p_{n}\right).\label{slope}
\end{equation}

Since motion in the gravitational field is a well-known problem, the
fundamental question in determining the dynamic evolution of the particle
will be to find the points of collision with the ground. To proceed
with this determination, we define the following two functions
\begin{equation}
\begin{split}G_{X}(p,t) & =x\left(p\right)-\left[x\left(p_{n}\right)+v_{x_{n}}^{(r)}t\right]\\
G_{Y}(p,t) & =y\left(p\right)-\left[y\left(p_{n}\right)+v_{y_{n}}^{(r)}t-\frac{g}{2}t^{2}\right],
\end{split}
\label{colideXY}
\end{equation}
where $\left(v_{x_{n}}^{(r)},v_{y_{n}}^{(r)}\right)$is the velocity
of the particle after it collides at point $p_{n}$. The next point
$p_{n+1}$ and the travel time $\delta t_{n,n+1}=(t_{n+1}-t_{n})$
spent by the particle between $p_{n}$ and $p_{n+1}$ are obtained
by solving the system of transcendental equations
\begin{equation}
\left\{ \begin{split}G_{X}(p_{n+1},\delta t_{n,n+1}) & =0\\
G_{Y}(p_{n+1},\delta t_{n,n+1}) & =0.
\end{split}
\right.\label{system}
\end{equation}
 In such a way, if the particles make a trip with $N$ collisions,
the total time spent will be
\begin{equation}
t_{N}=\sum_{n=1}^{N}\delta t_{n-1,n}\;\textrm{with}\;t_{0}=0.\label{tmax}
\end{equation}

In our model, we assume that only the component of the velocity normal
to the surface at the collision point is altered (inverted) \citep{Barreiro1}.
Then, at the instant of collision, the law of reflection relating
the incident velocity vector $\boldsymbol{v}_{n}^{(i)}$ to the reflected
velocity vector $\boldsymbol{v}_{n}^{(r)}$ is,
\begin{equation}
\boldsymbol{v}_{n}^{(r)}=\left(\boldsymbol{v}_{n}^{(i)}\cdot\hat{\boldsymbol{t}}_{n}\right)\hat{\boldsymbol{t}}_{n}-\gamma_{n}\left(\boldsymbol{v}_{n}^{(i)}\cdot\hat{\boldsymbol{n}}_{n}\right)\hat{\boldsymbol{n}}_{n}.\label{map1}
\end{equation}
{\small{}Obviously, the velocity vector, incident at a point $p_{n+1}$,
is related to the velocity vector reflected at the previous point
$p_{n}$ as
\[
\boldsymbol{v}_{n+1}^{(i)}=v_{x_{n}}^{(r)}\boldsymbol{i}+\left(v_{y_{n}}^{(r)}-g\,\delta t_{n,n+1}\right)\boldsymbol{j}.
\]
Now we can define the following dimensionless variables $\bar{x}(p)=x(p)/gt_{N}^{2}$,
$\bar{y}(p)=y(p)/gt_{N}^{2},\bar{\boldsymbol{v}}_{n}^{(r)}=\boldsymbol{v}_{n}^{(r)}/gt_{N}$
and $\phi_{n}=t_{n}/t_{N}$ , where $t_{N}$ is defined in (\ref{tmax}).
Therefore, the dimensionless velocity vector components in (\ref{map1})
take the form
\begin{equation}
\begin{split}\bar{v}_{x_{n+1}}^{(r)}= & \frac{\left(1-\gamma_{n+1}\lambda_{n+1}^{2}\right)\bar{v}_{x_{n}}^{(r)}+\lambda_{n+1}\left(1+\gamma_{n+1}\right)\left(\bar{v}_{y_{n}}^{(r)}-\delta\phi_{n,n+1}\right)}{1+\lambda_{n+1}^{2}}\\
\bar{v}_{y_{n+1}}^{(r)}= & \frac{\lambda_{n+1}\left(1+\gamma_{n+1}\right)\bar{v}_{x_{n}}^{(r)}+\left(\lambda_{n+1}^{2}-\gamma_{n+1}\right)\left(\bar{v}_{y_{n}}^{(r)}-\delta\phi_{n,n+1}\right)}{1+\lambda_{n+1}^{2}}.
\end{split}
\label{map2}
\end{equation}
and the system (\ref{system})} becomes
\begin{equation}
\left\{ \begin{split} & p_{n+1}=p_{n}+\frac{\bar{v}_{x_{n}}^{(r)}}{\bar{\alpha}}\delta\phi_{n,n+1}\\
 & \textrm{cos}\left(p_{n+1}\right)=\textrm{cos}\left(p_{n}\right)+\frac{\bar{v}_{y_{n}}^{(r)}}{\bar{\beta}}\delta\phi_{n,n+1}-\frac{1}{2\bar{\beta}}\delta\phi_{n,n+1}^{2}
\end{split}
\right.\label{system2}
\end{equation}
where $\bar{\alpha}=\alpha/gt_{N}^{2}$ and $\bar{\beta}=\beta/gt_{N}^{2}$.
Given the values of $p_{n}$, $\bar{v}_{x_{n}}^{(r)}$ and $\bar{v}_{y_{n}}^{(r)}$
of the $n$th iteration, the set of equations (\ref{system2}) produce
the values of $p_{n+1}$ and the travel time $\delta\phi_{n,n+1}$
which allows us to find $\bar{v}_{x_{n+1}}^{(r)}$ and $\bar{v}_{y_{n+1}}^{(r)}$through
(\ref{map2}). After that, the iterative process restart. 

\subsection{Conservative case}

We shall only consider the conservative scenario, when $\gamma_{n}$=$\gamma_{n+1}=1$.
Since we choose $\bar{\beta}\ll1$, it is appropriate to consider
that the point of collision with the ground has a height $\bar{y}(p_{n})\simeq\bar{y}(p_{n+1})\simeq0$
, but with local slope not necessarily zero. This approach avoids
transcendental equations and simplifies the calculation. As a consequence,
the second of the equations in (\ref{system2}) yields $\delta\phi_{n,n+1}=\phi_{n,n+1}-\phi_{n}=2\bar{v}_{y_{n}}^{(r)}$
. Finally, a simplified form of the map equations used to explain
motion is expressed as{\small{}
\begin{equation}
\begin{split}\bar{v}_{x_{n+1}}^{(r)}= & F_{1}\left(\bar{v}_{x_{n}}^{(r)},\bar{v}_{y_{n}}^{(r)},p_{n}\right)\\
\bar{v}_{y_{n+1}}^{(r)}= & \left|F_{2}\left(\bar{v}_{x_{n}}^{(r)},\bar{v}_{y_{n}}^{(r)},p_{n}\right)\right|\\
p_{n+1}= & F_{3}\left(\bar{v}_{x_{n}}^{(r)},\bar{v}_{y_{n}}^{(r)},p_{n}\right)
\end{split}
\label{map3-1}
\end{equation}
where
\begin{equation}
F_{1}\left(\bar{v}_{x_{n}}^{(r)},\bar{v}_{y_{n}}^{(r)},p_{n}\right)=\frac{\left(1-\bar{\lambda}_{n}^{2}\right)\bar{v}_{x_{n}}^{(r)}-2\bar{\lambda}_{n}\bar{v}_{y_{n}}^{(r)}}{1+\bar{\lambda}_{n}^{2}}
\end{equation}
\begin{equation}
\begin{split}F_{2}\left(\bar{v}_{x_{n}}^{(r)},\bar{v}_{y_{n}}^{(r)},p_{n}\right)= & \frac{2\bar{\lambda}_{n}\bar{v}_{x_{n}}^{(r)}+\left(1-\bar{\lambda}_{n}^{2}\right)\bar{v}_{y_{n}}^{(r)}}{1+\bar{\lambda}_{n}^{2}}\end{split}
\label{map3-2}
\end{equation}
\begin{equation}
F_{3}\left(\bar{v}_{x_{n}}^{(r)},\bar{v}_{y_{n}}^{(r)},p_{n}\right)=p_{n}+\frac{2}{\alpha}\bar{v}_{x_{n}}^{(r)}\bar{v}_{y_{n}}^{(r)}
\end{equation}
and were defined}{\footnotesize{}
\begin{equation}
\begin{split}\bar{\lambda}_{n}=\lambda_{n+1}= & -\frac{\bar{\beta}}{\bar{\alpha}}\textrm{sin}\left(p_{n}+\frac{2}{\alpha}\bar{v}_{x_{n}}^{(r)}\bar{v}_{y_{n}}^{(r)}\right).\end{split}
\label{map3-3}
\end{equation}
}{\footnotesize\par}

The ground was assumed to be flat, as a consequence there is a possibility
that $\bar{v}_{y_{n+1}}^{(r)}=F_{2}<0$. This non-physical situation
is bypassed by introducing the modulus in the second equation of (\ref{map3-1}).
This means that if such a case happens, the particle is re-injected
back to the dynamics with the same velocity but with a positive direction.

\subsubsection{Jacobian Matrix}

The Jacobian matrix $J=\partial\left(F_{1},F_{2},F_{3}\right)/\partial\left(v_{x},v_{y},p\right)$
for this dynamical system may be simply calculated using equations
(\ref{map3-1}-\ref{map3-3}), leading to \footnote{In Jacobian expressions, we utilize $(v_{x},v_{y},p)$ rather than
$(\bar{v}_{x_{n}}^{(r)},\bar{v}_{y_{n}}^{(r)},p_{n})$ to simplify
notation.}
\begin{widetext}
{\scriptsize{}
\begin{eqnarray*}
\frac{\partial F_{1}}{\partial v_{x}} & = & \frac{\bar{\alpha}^{4}+4\bar{\beta}\text{\ensuremath{v_{y}}}^{2}\cos\left(p+\frac{2\text{\ensuremath{v_{x}}}\text{\ensuremath{v_{y}}}}{\bar{\alpha}}\right)\left(\bar{\alpha}^{2}-\bar{\beta}^{2}\sin^{2}\left(p+\frac{2\text{\ensuremath{v_{x}}}\text{\ensuremath{v_{y}}}}{\bar{\alpha}}\right)\right)-\bar{\beta}^{4}\sin^{4}\left(p+\frac{2\text{\ensuremath{v_{x}}}\text{\ensuremath{v_{y}}}}{\bar{\alpha}}\right)-4\bar{\alpha}\bar{\beta}^{2}\text{\ensuremath{v_{x}}}\text{\ensuremath{v_{y}}}\sin\left(2p+\frac{4\text{\ensuremath{v_{x}}}\text{\ensuremath{v_{y}}}}{\bar{\alpha}}\right)}{\left(\bar{\alpha}^{2}+\bar{\beta}^{2}\sin^{2}\left(p+\frac{2\text{\ensuremath{v_{x}}}\text{\ensuremath{v_{y}}}}{\bar{\alpha}}\right)\right)^{2}}\\
\frac{\partial F_{1}}{\partial v_{y}} & = & \frac{2\bar{\beta}\left(\bar{\alpha}\left(-2\bar{\beta}\text{\ensuremath{v_{x}}}^{2}\sin\left(2p+\frac{4\text{\ensuremath{v_{x}}}\text{\ensuremath{v_{y}}}}{\bar{\alpha}}\right)+\bar{\alpha}^{2}\sin\left(p+\frac{2\text{\ensuremath{v_{x}}}\text{\ensuremath{v_{y}}}}{\bar{\alpha}}\right)+\bar{\beta}^{2}\sin^{3}\left(p+\frac{2\text{\ensuremath{v_{x}}}\text{\ensuremath{v_{y}}}}{\bar{\alpha}}\right)\right)+2\text{\ensuremath{v_{x}}}\text{\ensuremath{v_{y}}}\cos\left(p+\frac{2\text{\ensuremath{v_{x}}}\text{\ensuremath{v_{y}}}}{\bar{\alpha}}\right)\left(\bar{\alpha}^{2}-\bar{\beta}^{2}\sin^{2}\left(p+\frac{2\text{\ensuremath{v_{x}}}\text{\ensuremath{v_{y}}}}{\bar{\alpha}}\right)\right)\right)}{\left(\bar{\alpha}^{2}+\bar{\beta}^{2}\sin^{2}\left(p+\frac{2\text{\ensuremath{v_{x}}}\text{\ensuremath{v_{y}}}}{\bar{\alpha}}\right)\right)^{2}}\\
\frac{\partial F_{1}}{\partial p} & = & \frac{2\bar{\alpha}\bar{\beta}\cos\left(p+\frac{2\text{\ensuremath{v_{x}}}\text{\ensuremath{v_{y}}}}{\bar{\alpha}}\right)\left(\bar{\alpha}^{2}\text{\ensuremath{v_{y}}}-\bar{\beta}\sin\left(p+\frac{2\text{\ensuremath{v_{x}}}\text{\ensuremath{v_{y}}}}{\bar{\alpha}}\right)\left(\bar{\beta}\text{\ensuremath{v_{y}}}\sin\left(p+\frac{2\text{\ensuremath{v_{x}}}\text{\ensuremath{v_{y}}}}{\bar{\alpha}}\right)+2\bar{\alpha}\text{\ensuremath{v_{x}}}\right)\right)}{\left(\bar{\alpha}^{2}+\bar{\beta}^{2}\sin^{2}\left(p+\frac{2\text{\ensuremath{v_{x}}}\text{\ensuremath{v_{y}}}}{\bar{\alpha}}\right)\right)^{2}}
\end{eqnarray*}
}{\scriptsize\par}

{\scriptsize{}
\begin{eqnarray*}
\frac{\partial F_{2}}{\partial v_{x}} & = & -\frac{2\bar{\beta}\left(\bar{\alpha}\left(2\bar{\beta}\text{\ensuremath{v_{y}}}^{2}\sin\left(2p+\frac{4\text{\ensuremath{v_{x}}}\text{\ensuremath{v_{y}}}}{\bar{\alpha}}\right)+\bar{\alpha}^{2}\sin\left(p+\frac{2\text{\ensuremath{v_{x}}}\text{\ensuremath{v_{y}}}}{\bar{\alpha}}\right)+\bar{\beta}^{2}\sin^{3}\left(p+\frac{2\text{\ensuremath{v_{x}}}\text{\ensuremath{v_{y}}}}{\bar{\alpha}}\right)\right)+2\text{\ensuremath{v_{x}}}\text{\ensuremath{v_{y}}}\cos\left(p+\frac{2\text{\ensuremath{v_{x}}}\text{\ensuremath{v_{y}}}}{\bar{\alpha}}\right)\left(\bar{\alpha}^{2}-\bar{\beta}^{2}\sin^{2}\left(p+\frac{2\text{\ensuremath{v_{x}}}\text{\ensuremath{v_{y}}}}{\bar{\alpha}}\right)\right)\right)}{\left(\bar{\alpha}^{2}+\bar{\beta}^{2}\sin^{2}\left(p+\frac{2\text{\ensuremath{v_{x}}}\text{\ensuremath{v_{y}}}}{\bar{\alpha}}\right)\right)^{2}}\\
\frac{\partial F_{2}}{\partial v_{y}} & = & \frac{\bar{\alpha}^{4}-\bar{\beta}\left(4\text{\ensuremath{v_{x}}}^{2}\cos\left(p+\frac{2\text{\ensuremath{v_{x}}}\text{\ensuremath{v_{y}}}}{\bar{\alpha}}\right)\left(\bar{\alpha}^{2}-\bar{\beta}^{2}\sin^{2}\left(p+\frac{2\text{\ensuremath{v_{x}}}\text{\ensuremath{v_{y}}}}{\bar{\alpha}}\right)\right)+\bar{\beta}^{3}\sin^{4}\left(p+\frac{2\text{\ensuremath{v_{x}}}\text{\ensuremath{v_{y}}}}{\bar{\alpha}}\right)+4\bar{\alpha}\bar{\beta}\text{\ensuremath{v_{x}}}\text{\ensuremath{v_{y}}}\sin\left(2p+\frac{4\text{\ensuremath{v_{x}}}\text{\ensuremath{v_{y}}}}{\bar{\alpha}}\right)\right)}{\left(\bar{\alpha}^{2}+\bar{\beta}^{2}\sin^{2}\left(p+\frac{2\text{\ensuremath{v_{x}}}\text{\ensuremath{v_{y}}}}{\bar{\alpha}}\right)\right)^{2}}\\
\frac{\partial F_{2}}{\partial p} & = & -\frac{2\bar{\alpha}\bar{\beta}\cos\left(p+\frac{2\text{\ensuremath{v_{x}}}\text{\ensuremath{v_{y}}}}{\bar{\alpha}}\right)\left(\bar{\beta}\sin\left(p+\frac{2\text{\ensuremath{v_{x}}}\text{\ensuremath{v_{y}}}}{\bar{\alpha}}\right)\left(2\bar{\alpha}\text{\ensuremath{v_{y}}}-\bar{\beta}\text{\ensuremath{v_{x}}}\sin\left(p+\frac{2\text{\ensuremath{v_{x}}}\text{\ensuremath{v_{y}}}}{\bar{\alpha}}\right)\right)+\bar{\alpha}^{2}\text{\ensuremath{v_{x}}}\right)}{\left(\bar{\alpha}^{2}+\bar{\beta}^{2}\sin^{2}\left(p+\frac{2\text{\ensuremath{v_{x}}}\text{\ensuremath{v_{y}}}}{\bar{\alpha}}\right)\right)^{2}}\\
\frac{\partial F_{3}}{\partial v_{x}} & = & \frac{2\text{\ensuremath{v_{y}}}}{\bar{\alpha}}\\
\frac{\partial F_{3}}{\partial v_{y}} & = & \frac{2\text{\ensuremath{v_{x}}}}{\bar{\alpha}}\\
\frac{\partial F_{3}}{\partial p} & = & 1
\end{eqnarray*}
}{\scriptsize\par}
\end{widetext}

\LyXZeroWidthSpace It is straightforward to show that the Jacobian
matrix's determinant is equal to one, confirming that the system is
indeed conservative and the dimensionless energy 
\begin{equation}
\bar{E}_{n}=\frac{1}{2}\left[\left(\bar{v}_{x_{n}}^{(r)}\right)^{2}+\left(\bar{v}_{y_{n}}^{(r)}\right)^{2}\right]\label{Energy}
\end{equation}
is constant.

\subsection{Periodic points \label{subsec:Periodic-points}}

\begin{figure}[H]
\centering{}\includegraphics[scale=0.5]{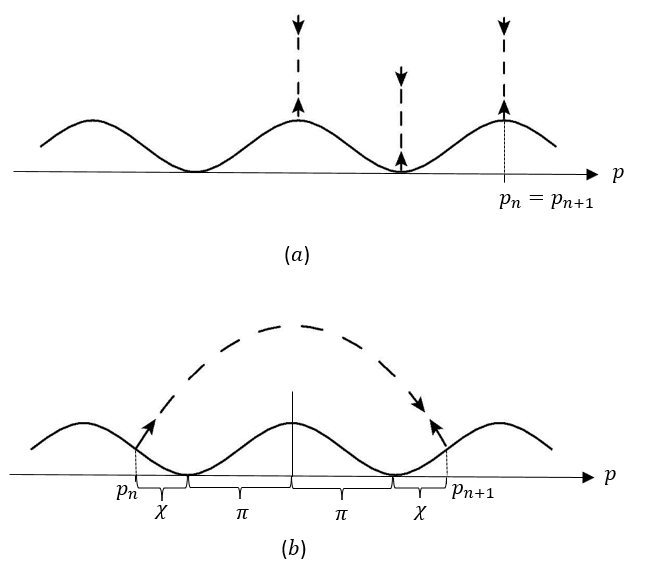}\caption{Examples of fixed points: (a) Fixed point of period one. The dynamical
system returns to the point in phase space at each iteration and (b)
the system returns to the point after 2 iterations.}
\label{FigFixedPoints}
\end{figure}

We can anticipate the occurrence of some exceptional points using
the physics of the problem. These are known as fixed points, to which
the dynamical system returns after one iteration (period-one fixed
point), two iterations (period-two fixed point), or n iterations (period-n
fixed point). The figure \ref{FigFixedPoints} illustrates two fixed
points: $(a)$ Fixed points for period one and $(b)$ Fixed points
for period two.

\subsubsection{Period-one Point}

It is evident that period-one fixed points, as shown in portion $(a)$
of the figure, have a zero local slope. So, as long as the x component
of the initial velocity is zero, the system will not experience any
diffusion in the horizontal axis. A period-one point is obtained by
solving the following equations: $\bar{v}_{x_{n+1}}^{(r)}=\bar{v}_{x_{n}}^{(r)}=0$,
$\bar{v}_{y_{n+1}}^{(r)}=\bar{v}_{y_{n}}^{(r)}$ and $p_{n+1}=p_{n}$
with $\bar{\lambda}_{n}=0$ (zero slope). We can verify the fact considering
first equation in (\ref{map3-3})
\[
\bar{\lambda}_{n}=0\Rightarrow\textrm{sin}\left(p_{n}+\frac{2}{\alpha}\bar{v}_{x_{n}}^{(r)}\bar{v}_{y_{n}}^{(r)}\right)=0\underset{\bar{v}_{x_{n}}^{(r)}=0}{\Rightarrow}p_{n}=m\pi,
\]
where $m$ is a integer. These points indicate the locations of the
peaks and valleys in Figure 1 - part (a). Thus
\begin{eqnarray}
\bar{v}_{x_{n+1}}^{(r)} & = & F_{1}\left(0,\bar{v}_{y_{n}}^{(r)},m\pi\right)=0\nonumber \\
\bar{v}_{y_{n+1}}^{(r)} & = & F_{2}\left(0,\bar{v}_{y_{n}}^{(r)},m\pi\right)=\bar{v}_{y_{n}}^{(r)}\label{peri1}\\
p_{n+1} & = & F_{3}\left(0,\bar{v}_{y_{n}}^{(r)},m\pi\right)=m\pi\nonumber 
\end{eqnarray}
We have the following physical situation: If a particle is chosen
whose horizontal component of velocity is zero, in a zero slope point,
clearly the $x$-coordinate of the particle will never change and
the particle does not scatter in the $x$-direction. 

\subsubsection{Period-two Points}

We now consider points with non-zero slope. In general, the particle
gains a non-zero horizontal component to the velocity and then diffuses
along the horizontal axis. Nevertheless, depending on the initial
conditions, it is possible for the particle to strike the surface
at point $p_{n}$ with velocity $\vec{v}_{n}$, reflect there, then
it reaches point $p_{n+1}$ with velocity $\vec{v}_{n+1}$, where
it will then reflect again and go back to point $p_{n}$ with velocity
$\vec{v}_{n}$. Part $(b)$ of Fig. \ref{FigFixedPoints} depicts
an illustration of this kind.. Inspired by the figure, consider points
connected by $\bar{v}_{x_{n+2}}^{(r)}=-\bar{v}_{x_{n+1}}^{(r)}=\bar{v}_{x_{n}}^{(r)}$,
$\bar{v}_{y_{n+2}}^{(r)}=\bar{v}_{y_{n+1}}^{(r)}=\bar{v}_{y_{n}}^{(r)}$,
$p_{n+2}=p_{n}$, $\bar{y}(p_{n+1})=\bar{y}(p_{n})$ and opposite
local slopes $\lambda_{n+1}=-\lambda_{n}$. 

Taking into account the figure \ref{FigFixedPoints} portion (b) ,
the points $p_{n}$ and $p_{n+1}$ must be connected by
\[
\textrm{ }\begin{cases}
p_{n}=-\pi-\chi\\
p_{n+1}=\pi+\chi
\end{cases}\textrm{with }0<\chi<\pi
\]
where we are solely concerned with the most straightforward solution.
Then, with the help of equations (\ref{system2}), we can write 
\begin{equation}
\bar{v}_{x_{n}}^{(r)}\bar{v}_{y_{n}}^{(r)}=\bar{\alpha}\left(\pi+\chi\right).\label{peri2A}
\end{equation}
In addition, the first of the equations (\ref{map3-2}) yields
\begin{equation}
\bar{v}_{y_{n}}^{(r)}=\frac{\bar{\alpha}}{\bar{\beta}\textrm{sin}\left(\chi\right)}\bar{v}_{x_{n}}^{(r)}.\label{peri2B}
\end{equation}
These results allow us to determine both $\bar{v}_{x_{n}}^{(r)}$
and $\bar{v}_{y_{n}}^{(r)}$ as functions of $\chi$. So the period
two fixed point is written as
\begin{eqnarray*}
\bar{v}_{x_{n}}^{(r)} & = & \pm\sqrt{\bar{\beta}\left(\pi+\chi\right)\textrm{sin}\left(\chi\right)}\\
\bar{v}_{y_{n}}^{(r)} & = & \frac{\bar{\alpha}\left(\pi+\chi\right)}{\sqrt{\bar{\beta}\left(\pi+\chi\right)\textrm{sin}\left(\chi\right)}},\\
p_{n} & = & \mp\left(\pi+\chi\right)
\end{eqnarray*}

\begin{figure}[H]
\centering{}\includegraphics[scale=0.4]{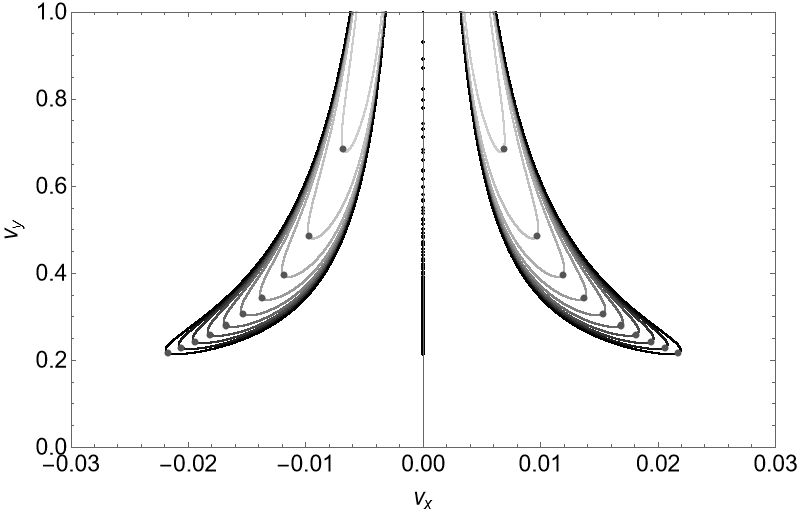}\caption{Period one fixed points are represented by the black dots in the center
of the line. The other points are the period 2 fixed points.The gray
dots at the end of the curves are the points obtained with the value
$\chi=\pi/2.$}
\label{FigPeriod1and2}
\end{figure}

Figure \ref{FigPeriod1and2} illustrates these fixed points. The middle
points in black in this picture indicate the period-1 fixed points.
The graphic also illustrates the effect of the $\beta-$parameter
on the formation of period-2 fixed points. The points are calculated
by altering the value of $\chi$ from 0 to $\pi$, and each gray level
indicates a $\beta$ parameter value from the lightest gray $(\beta=0.00001)$
to the darkest $(\beta=0.0001)$. $\alpha=0.001$ is used for all
points. 

The choice $\chi=\pi/2$ is used to calculate the gray dots in the
figure \ref{FigPeriod1and2}. Each curve is divided into two branches
by these points. The points that make up the branches we name external
have $\chi>\pi/2$, whereas the points that make up the branches we
term internal have $\chi<\pi/2$ . Consider the eigenvalues of the
Jacobian matrix to categorize the stability of these points. The external
points $\left(\chi\geq\pi/2\right)$ can be classified as node-type
stable points since the modules of their Jacobian matrix eigenvalues
are all equal to 1. On the other hand, because all of the eigenvalues
are real with one positive and the others negatives, the internal
points $\left(\chi<\pi/2\right)$ are categorized as unstable points
of the saddle type. Therefore, the gray dots in the phase space represent
saddle-node bifurcations \citep{Kuznetsov1995}.

Many more types of fixed points may exist, and this subject will be
addressed in future work \citep{Barreiro2}. We are mostly interested
in the particle dispersion problem along the horizontal axis in this
work.

\section{Diffusion Process}

\subsection{The stochastic character of force}

\begin{figure}[h]
\centering{}\includegraphics[scale=0.33]{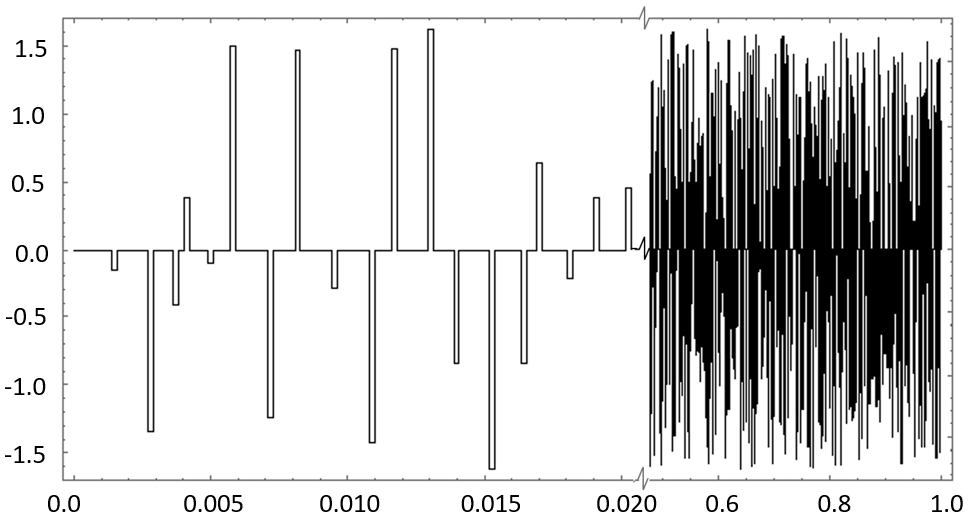}\caption{Typical behavior of the horizontal component of the collision force.
Here we have used $\bar{\alpha}=0.01$ and $\bar{\beta}=0.005$. The
graph has two regions with different scales. On the left we have the
region magnified between $\phi=0.000$ and $\phi=0.020$ and on the
right, after a cut in the graph, the normal scale from $\phi=0.5$
to $\phi=1.0$ is shown.}
\label{fig:Force}
\end{figure}

Clearly, unless we are in some special initial point, the particles
must diffuse in the $x$-direction. This diffusion is caused by the
collision force with the ground. Due to the irregular nature of the
ground, the collision force $\boldsymbol{F}_{col}$ has components
in both horizontal and vertical directions. It is intuitive to notice
that the horizontal component presents different magnitudes and directions
at each collision. To understand the behavior of this horizontal component
of the collision force, we can describe it as
\[
\bar{F}_{col_{x}}(\phi_{n})=\left.\frac{\Delta\bar{v}}{\bar{\tau}}\right|_{p_{n}}=\frac{\bar{v}_{x_{n}}^{(r)}-\bar{v}_{x_{n}}^{(i)}}{\bar{\tau}}=\frac{\bar{v}_{x_{n}}^{(r)}-\bar{v}_{x_{n-1}}^{(r)}}{\bar{\tau}}
\]
where $\bar{\tau}$ is the dimensionless collision time, which is
extremely small. We will also assume that the collision force is approximately
constant during the collision time and a typical example of what this
force looks like is shown in figure \ref{fig:Force}.

The width of each rectangle represents the collision time and despite
the dynamics being well known and the irregularities in the ground
having a periodicity, the numerical results presented show that the
effects of the horizontal component of this force has a behavior comparable
to a stochastic force. It is actually extremely difficult to tell
whether a sequence is random or chaotic, but there are some proposed
procedures to distinguish between these two behaviors. In this work
we will make use of the permutation entropy (PE) method \citep{Bandt,Riedl2013}
to establish the randomness of the time series produced by the collision
force. Denoting the time series as $\{S_{t}\}_{t=1,\ldots,T}$ the
method consists in defining subsets of order $\mathcal{O}$, forming
the set $S=\{\{S_{1},S_{2},\ldots,S_{\mathcal{O}}\},\{S_{2},S_{3},\ldots,S_{\mathcal{O}+1}\},$
$\ldots,\{S_{T-\mathcal{O}+1},\ldots,$ $S_{T-1},S_{T}\}\}$. We then
compare consecutive values from each subset to establish the associated
permutation. For example, $\{S_{1}<S_{2}<\ldots<S_{\mathcal{O}}\}$
represents the permutation $\{1,2,...,O\}$, while $\{S_{2}<S_{1}<\ldots<S_{\mathcal{O}}\}$
represents the permutation ${2,1,...,O}$ and so on, yielding the
set of all permutations associated with the sequence $S$, named $\varPi(S)$.
Then, the set of all $O!$ possible permutations $\pi_{i}$ of the
numbers $\{1,2,...,O\}$ are constructed. The relative frequency of
each permutation $\pi_{i}$ can be calculated by counting the number
of times the permutation $\pi_{i}$ is found in the set $\varPi(S)$
divided by the total number of sequences,
\begin{equation}
P_{i}=\frac{\textrm{Number of times that \ensuremath{\pi_{i}} appears in \ensuremath{\varPi}(S) }}{T-\mathcal{O}+1}.\label{probabil}
\end{equation}
and the normalized permutation entropy function is written as, 

\begin{equation}
PE_{\mathcal{O}}=-\frac{1}{\log_{2}(\mathcal{O}!)}\sum_{i=1}^{\mathcal{O}!}P_{i}\log_{2}(P_{i}).\label{entropy}
\end{equation}

Formulas (\ref{probabil}) and (\ref{entropy}) were applied to the
temporal sequences of collision forces for three different initial
conditions and also different orders $\mathcal{O}$. The table \ref{table}
shows the results obtained. 
\begin{table*}
\centering{}%
\begin{tabular}{|c|c|c|c|c|c|}
\hline 
floor parameters & initial condition & $\mathcal{O}=3$ & $\mathcal{O}=4$ & $\mathcal{O}=5$ & $\mathcal{O}=6$\tabularnewline
\hline 
\hline 
\multirow{2}{*}{$\begin{array}{c}
\alpha=0.01\\
\beta=0.05
\end{array}$} & $p_{0}=-0.033$ & \multirow{1}{*}{0.998569} & \multirow{1}{*}{0.995189} & \multirow{1}{*}{0.981222} & \multirow{1}{*}{0.92671}\tabularnewline
\cline{2-6} \cline{3-6} \cline{4-6} \cline{5-6} \cline{6-6} 
 & $p_{0}=0.032$ & \multirow{1}{*}{0.999633} & 0.995120 & 0.982245 & 0.925946\tabularnewline
\hline 
\multirow{2}{*}{$\begin{array}{c}
\alpha=0.01\\
\beta=0.0005
\end{array}$} & $p_{0}=-0.033$ & 0.998874 & 0.994082 & 0.986440 & 0.935262\tabularnewline
\cline{2-6} \cline{3-6} \cline{4-6} \cline{5-6} \cline{6-6} 
 & $p_{0}=0.032$ & 0.999501 & 0.996295 & 0.984878 & 0.934281\tabularnewline
\hline 
\end{tabular}\caption{The initial conditions are chosen in order to vary the initial point
$(x(p_{0}),y(p_{0}))$ and keeping the energy $\bar{E}=4$ constant.}
\label{table}
\end{table*}
The smaller the $PE_{\mathcal{O}}$ is, the more regular and more
deterministic the time series is. Contrarily, the closer to 1 the
value of $PE_{\mathcal{O}}$ is, the more noisy and random the time
series is. The results allow us to assume that the force is random.

\section{Probability Distribution Function (PDF)}

\begin{figure*}
\begin{centering}
\includegraphics[scale=0.6]{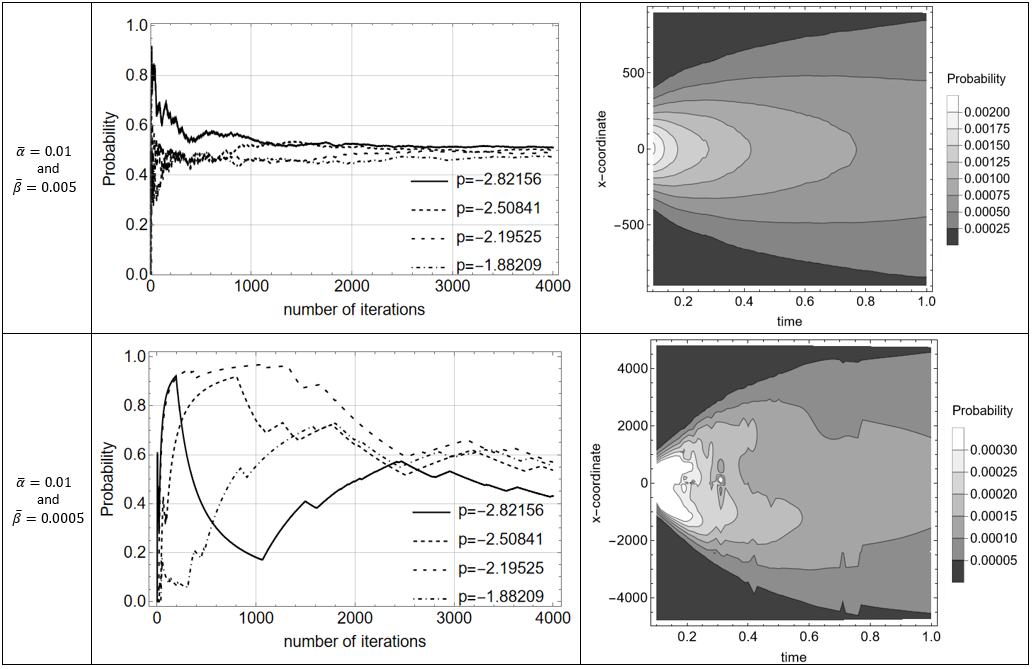}
\par\end{centering}
\caption{The first graphic of each column contains time evolution examples
for the likelihood of a single particle jumping to the right. The
difference is in the $\beta$ parameter value, which is lowered to
one-tenth and one-hundredth of its initial value in the columns on
the left. The evolution of the 4-particle leaps (4 initial conditions)
is explored in the graphs. The different initial conditions for the
particles are obtained by changing the initial parameter $p$ in the
functions $x(p)$ and $y(p)$ in Eq (\ref{xyspt}) and keeping the
energy $\bar{E}=4$ constant. The selected $p$-parameters are shown
in the figures. The respective contour plots for the probability distributions
are shown on the right.}
\label{distribProb}
\end{figure*}

This section's major purpose is to establish the probability distribution
function (PDF) $\Psi(x,t)$, which provides us the probability of
the particle being on the coordinate $x$ at time $t$, and what it
has to do with normal and superdiffusive processes. Among the various
diffusive processes, Brownian motion is the prototype for the description
of non-equilibrium dynamical systems. Due to the stochastic behavior
of the collision force, the jumps performed by the particles also
reproduce characteristics of random walk. We can comprehend this by
calculating the chance of each particle going to the right. After
each impact, we obtain the $x-$component of the velocity. Then, by
examining the sign of these velocities and associating $+1$ for $v_{x}>0$
and $0$ for $v_{x}<0$, we can count the number of jumps to the right
and derive the evolution of this probability as the number of jumps
increases. It is appropriate at this point to introduce an index that
specifies the initial condition ($\nu$), which is used to compute
the Probability Density Function (PDF) for the complete ensemble.
So, starting with an initial state labeled by $\nu$, the probability
of jumping to the right after $n$ jumps is calculated as follows:
\begin{eqnarray*}
P_{r-jump}(n,\nu) & = & \frac{1}{n}\sum_{i=1}^{n}SgnPlus(v_{x,i}^{(\nu)})\\
 &  & \textrm{ where }SgnPlus(v_{x,i}^{(\nu)})=\begin{cases}
1 & \textrm{if }v_{x}>0\\
0 & \textrm{if }v_{x}<0
\end{cases}
\end{eqnarray*}
Figure \ref{distribProb}, on the left, shows examples of the time
progression of individual particle jumps for four distinct initial
conditions and two ground parameter adjustments, as well as the corresponding
PDFs $\Psi(x,t)$. With time evolution, the left/right jump probabilities
for a ground with $\alpha=0.01$ and $\beta=0.005$ tend to be 0.5
very quickly as we can see into upper graphic on the left. However,
if the beta parameter is set to $\beta=0.0005$ the graph indicates
an initial oscillation, but the probability ultimately tends to reach
0.5. 

The coordinates of the collision points and the travel time between
one point and the next are obtained from the mapping given in equations
(\ref{map2}) and (\ref{system2}). It is obvious that the travel
time varies between jumps. However, for our analysis, it is critical
to obtain the particle's position as a function of time with equal
time intervals. This is simple because the particle moves in a gravitational
field $\boldsymbol{g}$, and we can easily calculate its position
as a function of time. The time is then normalized so that the maximum
time equals one. So, to get the probability distribution, for all
iterative processes, we begin by subtracting the starting position
of the particles. As a result, all of the particles in the ensemble
start from the same position. In our scenario, we have 2000 particles
performing 4000 leaps, totaling 8 million collision points, but it
is clear that the number of points as a function of time depends on
the choice of interval $dt$ and can be much higher. To demonstrate
the procedure, the simulation is configured so that each particle
in the ensemble has an energy of $E=4$. The outcomes for two different
types of grounds are shown in Figure \ref{distribProb} on the right.

The first PDF graph was obtained with the parameters $\alpha=0.01$
and $\beta=0.005$, and shows a probability density region following
a format very similar to a Gaussian distribution. The second PDF,
obtained with the parameters $\alpha=0.01$ and $\beta=0.0005$, has
an extremely anomalous diffusion in the early part of its time evolution,
however when the time evolution takes place, the PDF apparently starts
to show a Gaussian behavior. In order to have a better understanding
of this behavior, we studied the moments associated with each distribution.
Inspired by the Gaussian form of normal diffusion, with an anomalous
diffusion we make a scaling hypothesis \citep{Cecconi2022} so that
we can express the anomalous distribution as
\begin{equation}
\Psi_{\mu}(x,t)=\sqrt{\frac{a}{\pi}}\frac{1}{t^{\mu}}\exp\left[-a\left(\frac{x}{t^{\mu}}\right)^{2}\right].\label{GaussForm}
\end{equation}
The associated moments are easily obtained as 
\begin{equation}
\left\langle \left|x(t)\right|^{m}\right\rangle =\intop_{-\infty}^{\infty}x^{m}\Psi_{\mu}(x,t)\,dx=\frac{1}{\sqrt{a^{m}\pi}}\varGamma\left(\frac{m+1}{2}\right)t^{m\mu}.\label{moments}
\end{equation}

\begin{figure*}
\centering{}\includegraphics[scale=0.6]{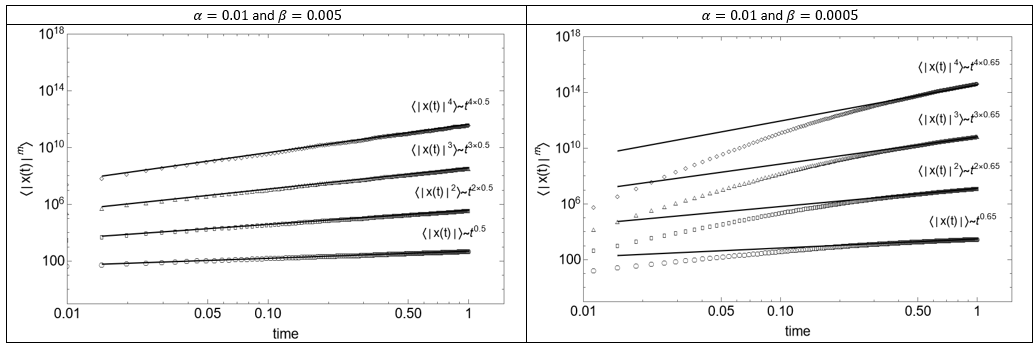}\caption{The black lines represent equations of powers $t^{m\mu}$. At left
we have a normal diffusion and at right we have a superdiffusive process.
We can see that more than half of the time evolution has passed before
the superdiffusive behavior with scale $\mu=0.65$ manifests.}
\label{figMSDa}
\end{figure*}

The result shows a behavior of MSD as $\left\langle x^{2}\right\rangle \propto t^{2\mu}$,
therefore normal distribution has a scale parameter $\mu=1/2$. If
$\mu<1/2$ we have a subdiffusive process and for $\mu>1/2$ we found
a superdiffusive behavior. Figure \ref{figMSDa} shows the results
of the moments calculations for two different grounds. We can observe
that at left we obtain the scale $\mu=0.5$ and at right we obtain
$\mu=0.65$. So, we have two distinct behaviors: at left a normal
diffusion and at right we have a superdiffusive behavior.

\begin{figure}
\centering{}\includegraphics[scale=0.38]{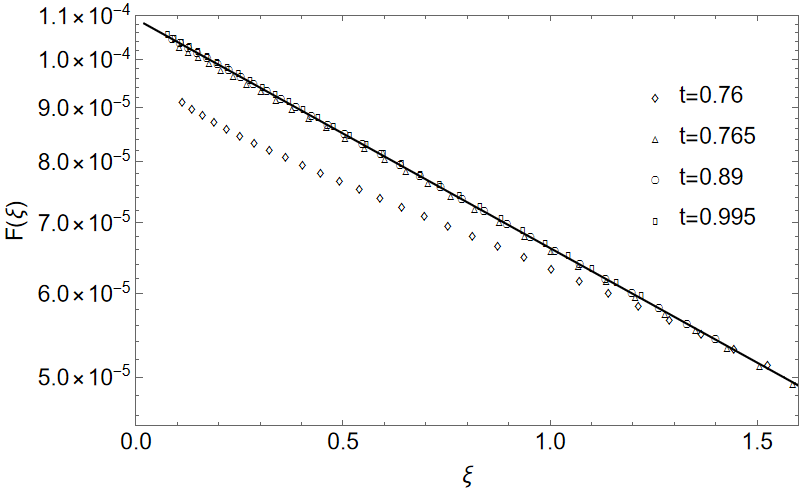}\caption{PDF's rescaled by the factor $\xi=x^{2}/\left\langle x(t)^{2}\right\rangle $for
four distinct times. The theoretical prediction given in Eq. (\ref{Fqui})
with an $a=3.75\times10^{-8}$ is shown by the black line. \label{fig:PDF-rescaled}}
\end{figure}

The scaling hypothesis is carried forward using equation (\ref{moments})
to obtain $t^{2\mu}=a\sqrt{\pi}\left\langle x(t)^{2}\right\rangle /\varGamma\left(3/2\right)$,
which enables us to specify the subsequent function
\begin{equation}
F(\xi)=t^{\mu}\Psi_{\mu}(x,t)=\sqrt{\frac{a}{\pi}}\exp\left[-\frac{\varGamma\left(3/2\right)}{\sqrt{\pi}}\xi\right]\label{Fqui}
\end{equation}
where $\xi=x^{2}/\left\langle x(t)^{2}\right\rangle $. Using the
PDF data for the superdiffusive process ($\mu=0.65$) we obtain $F(\xi)$
numerically and the results for $t=0.76$, $t=0.765$, $t=0.89$ and
$t=0.995$ are presented in the figure \ref{fig:PDF-rescaled} .
 The only parameter that can be adjusted in the theoretical forecast
stated in Eq (\ref{Fqui}) is the value of $a$. We get a remarkable
agreement with the simulation findings when we choose $a=3.75\times10^{-8}$.
The black dot-dashed line on the graph denotes the theoretical result
obtained in equation (\ref{Fqui}). We observe that the theoretical
modeling and the simulation outcome start to diverge for periods of
time less than 76.5\% of the overall duration of the iterative procedure.
Rescaling the data, all simulation points for times more than this
amount lie exactly on the same curve. This was already a foregone
conclusion if we look at the second PDF in the figure \ref{distribProb},
which shows quite anomalous behavior for times less than 0.8. Before
this time has elapsed the particles display a strongly anomalous diffusion
with a scale that must rely on the moment being estimated, $\left\langle \left|x\right|^{m}\right\rangle \propto t^{m\mu(m)}$,
\citep{Andersen2000}. 

\section{Conclusions and Outlook}

In this work, we have studied a falling particle in the gravitational
field colliding with a non-plane surface. We could observe that the
horizontal component of the collision force presented a stochastic
behavior. This was verified by using the entropy permutation method
applied to the collision force time series. Additionally, we established
that the jumps to the right and left follow a distribution whose probabilities
tend toward 0.5 while the particle's temporal development takes place.
It can be seen that the convergence to the factor 0.5 occurs significantly
more quickly using the ground with parameters $\alpha=0.01$ and $\beta=0.005$
than with $\beta=0.0005$. We assume that a surface with more pronounced
undulations produces a horizontal component of the force that swiftly
alters the particle's horizontal motion, causing the probability of
jumps to fast converge to 0.5. The first case implied a diffusion
process that follows Einstein's famous relationship so that the horizontal
mean square displacement is proportional to time, $\left\langle x(t)^{2}\right\rangle \sim t$.
The system begins to become superdiffusive as the ground gets smoother.
In fact, it is observed that the system with $\beta=0.0005$ exhibits
a strongly anomalous mean squared deviation with temporal increase
over the earliest portion of its temporal history. Subsequently, the
movement becomes \textquotedbl standard superdiffusive\textquotedbl .
To comprehend this behavior, we assumed that the probability density's
functional form must take on a Gaussian form of normal diffusion,
with the exception that the distribution's time dependence is scaled
by $t^{\mu}$. We get a remarkable consistency between the theoretical
expression and the simulation results using this approach. Future
works are being developed including changes in the function that describes
the floor, introduction of dissipation and oscillations in the ground,
among other works.

\section*{Acknowledgments}

The authors would like to thank Prof. Edson Denis Leonel for the observations
and comments, as well as Coordination for the Improvement of Higher
Education Personnel (Capes) for the financial support.

\section*{\textemdash \textemdash \textemdash \textemdash \textemdash \textendash{}}

\bibliographystyle{apsrev4-2}
\addcontentsline{toc}{section}{\refname}\bibliography{paper0Rev}

\end{document}